\documentclass[%
reprint,
amsmath,amssymb,
aps,
prx,
]{revtex4-2}

\usepackage{graphicx}
\usepackage{dcolumn}
\usepackage{bm}

\begin{document}
	
	\preprint{APS/123-QED}
	
	\title{Avalanche Dynamics in Stick-Slip Cutting of Molybdenum Disulfide}
	
	\author{Paweł Koczanowski}
	\affiliation{%
		Marian Smoluchowski Institute of Physics, Jagiellonian University, Kraków, Poland
	}%

	\author{Paolo Nicolini}

	\affiliation{Institute of Physics, Czech Academy of Sciences, Prague, Czech Republic
	}%

	\author{Hesam Khaksar}
	\affiliation{%
		Marian Smoluchowski Institute of Physics, Jagiellonian University, Kraków, Poland
	}%

	\author{Enrico Gnecco}
	\affiliation{%
		Marian Smoluchowski Institute of Physics, Jagiellonian University, Kraków, Poland
	}%

	\begin{abstract}
We have investigated nanoscale wear on multilayered MoS$_2$, the flagship transition metal dichalcogenide, by elastically driving sharp diamond tips under normal loads sufficient to induce in-plane fracture. The accompanying friction and the resulting wear structures were first characterized by atomic force microscopy (AFM), revealing a stick-slip regime that drives progressive exfoliation of MoS$_2$ chips. At high normal forces, the slip phase displays hallmark signatures of avalanche dynamics, observed for the first time at the nanoscale, evidenced by a Generalized Extreme Value distribution of friction force drops. The AFM characterization is corroborated by molecular dynamics simulations, which reproduce experimental trends and uncover atomistic details of the wear process, including local amorphization, layer curving, and the involvement of distinct dissipative channels. Notably, it appears that only one fifth of the energy inputted into the system is used to damage the MoS$_2$ surface irreversibly. These results offer new insight into the physical mechanisms governing friction and wear in layered solids and provide a framework for precision cutting and nanomachining in van der Waals materials, relevant to next-generation devices at sub-micrometer scales.
	\end{abstract}
	
	\maketitle
	
	
\section{Introduction}
	Our understanding of sliding friction at the nanoscale has made giant progress since Mate \emph{et al.} were able to observe atomic-scale stick-slip in the driven motion of a tungsten tip on a graphite surface \cite{1}. After this pioneering work, it took a few years to interpret the stick-slip phenomenon with a simple mechanical model introduced by Prandtl in 1928 \cite{2}. Basically, if a point particle (representing the tip apex) is confined in a minimum of an energy landscape determined by the crystal structure of the surface, and is elastically driven along a given direction, the spring force $F$ increases with time until the equilibrium becomes unstable and the particle slips into a neighbour equilibrium location. The sudden displacement of the tip is revealed by a force drop $\Delta F$ accompanied by energy loss. Atomic-scale stick-slip has been reported on a variety of crystal surfaces in different environments, from ultra-high vacuum (UHV) \cite{3,4} to water \cite{5}, using atomic force microscopy (AFM) in the so-called contact mode. Several details not accessible experimentally, such as the influence of the tip shape on the time evolution of the friction force, have been also unveiled using molecular dynamics (MD) simulations \cite{6}. However, most of the experimental and theoretical characterizations of atomic-scale stick-slip reported so far share a common trait: the stick-slip is not accompanied by surface wear. It makes the results statistically reproducible and easier to interpret, but practical applications are often quite far from this ideal case.
	
	The study of atomic-scale stick-slip accompanying abrasive wear possibly started from an earlier work on alkali halide crystal surfaces in UHV by one of us.\cite{7} On a KBr(001) surface scraped by a sharp silicon tip the effect was observed with loading forces of only few nN. In the presence of wear the baseline of the spring force profile becomes tilted, which was attributed to the contact between tip and surface getting larger and progressively worn out. Atomic-scale wear accompanying stick-slip motion in UHV was also reported on metal surfaces (Cu and Au) by Gosvami \emph{et al.} \cite{8} In this case, the wear onset was attested by the friction profile becoming very irregular, possibly due to material transfer onto the probing tip. On layered-materials, stick-slip motion accompanying AFM-based nanoscratching was reported by Özogul \emph{et al.} on MoS$_2$ and WSe$_2$ \cite{9,10}. The characteristic periods were found to be much larger than the lattice constant of the crystal surface, but a satisfactory explanation of this observation was not found.
	
	MD simulations of crystal surfaces scratched by a nanotip can be traced back to the work by Belak \emph{et al.} on Cu and Si substrates \cite{11}. Stick-slip was reproduced by \emph{Fang et al.} on Au and Pt thin films \cite{12}, and crystalline Cu and Si surfaces were again in the focus of more recent and detailed work by Zhang \emph{et al.} \cite{13} and Li \emph{et al.} \cite{14} A noticeable computational effort aimed to reproduce plowing wear on 2D materials at the nanoscale was presented by Klemenz \emph{et al.} in 2014 for the case of a graphene layer coating a Pt(111) surface \cite{15}. For normal force values below 200 nN the authors observed regular stick-slip, accompanied by elastic deformation of the Pt substrate. At the threshold value, the graphene layer went suddenly ruptured, with subsequent (plastic) deformation of the metal substrate and the formation of a scratch groove.
	
	Like graphite, MoS$_2$ has been used as a solid lubricant for decades \cite{16}. The experimental verification of superlubricity in the early nineties for multilayer MoS$_2$ coatings in UHV conditions \cite{17} sparked a significant interest for it. Since then, a lot of effort has been devoted in order to understand (and possibly control) the complex phenomena taking place during sliding on this material, such as the formation of crystalline layers from molecular precursors18 or from amorphous MoS$_2$ \cite{19}. On scratched MoS$_2$ monolayers, MD simulations have focused on the role of grain boundaries \cite{20}, defects \cite{21} and wrinkles \cite{22}, or revealed the formation of dislocations in the substrate \cite{23}. Addressing a multilayer material removes any complexity due to the presence of a third body (the substrate), in addition to the scratch tool and the scraped surface.
	
	In this article, we present nanoscale investigations of early-stage wear mechanisms in MoS$_2$ combining AFM experiments with MD simulations. While our focus is on molybdenum disulfide, a prototypical transition metal dichalcogenide, the mechanisms uncovered here are expected to be broadly relevant to other layered van der Waals materials. The presentation is organized as follows. We have first scratched MoS$_2$ with normal force values in the $\mu$N range, recorded the time evolution of the lateral (friction) force while scratching and imaged the resulting wear structures using AFM. The stick-slip motion of the scratching tool (a sharp diamond tip) is also interpreted within a standard `block and spring' model and the statistical distribution of the force drops observed experimentally are extracted. MD simulations close the circle. Not only they confirm the stick-slip behavior, but also shed light on atomistic details and on the energetics of the exfoliation process. 
	
	\section{Results}
	
	\subsection{ Experimental Investigation of Nanoscale Plowing Wear}
	Stick-slip is observed in all AFM measurements, regardless of the normal force $F_N$ (between 0.6 and 1.8 $\mu$N). Figure 1(a) illustrates three lateral force (friction) profiles obtained with $F_N$ = 0.6 $\mu$N, 1.1 $\mu$N and 1.4 $\mu$N. In all cases, the period $\lambda$ of the stick-slip is in the order of 20 nm and well above the lattice constant of MoS$_2$ (3.16 \AA). The stick-slip profile becomes irregular when $F_N$ is increased. In Figure 1(b), we have plotted the key parameters characterizing this process (force drop $\Delta F$, critical force $F_c$, lateral stiffness $k$, slip length $x_\mathrm{slip}$ and slip period $\lambda$), as functions of $F_N$ for a fixed value of the pulling speed $v = 1.25$ $\mu$m/s. Figure 1(c) shows that the critical force Fc is proportional to $F_N$: $F_c = A F_N$, with a slope $A = 1.81 \pm 0.04$ corresponding to the static friction coefficient. The same can be said for the force drop $\Delta F = B F_N$, with $B = 0.101 \pm 0.001$ (Figure 1d). The lateral stiffness $k$ increases also proportionally to $F_N$ ($k = C F_N$), with a slope $C = (9.9 \pm 0.8)\times 10^6$ m$^{-1}$ (Figure 1e). Not so surprisingly, this trend is different from an elastic contact, where, for a round shaped tip apex, $k \propto F_N^{1/2}$ \cite{24}. In contrast with $F_c$, $\Delta F$ and $k$, the period $\lambda$ (Figure 1f) and the slip length $x_\mathrm{slip}$ (Figure 1h) are found not to increase with $F_N$.
	
	\begin{figure*}
		\includegraphics[width = \textwidth]{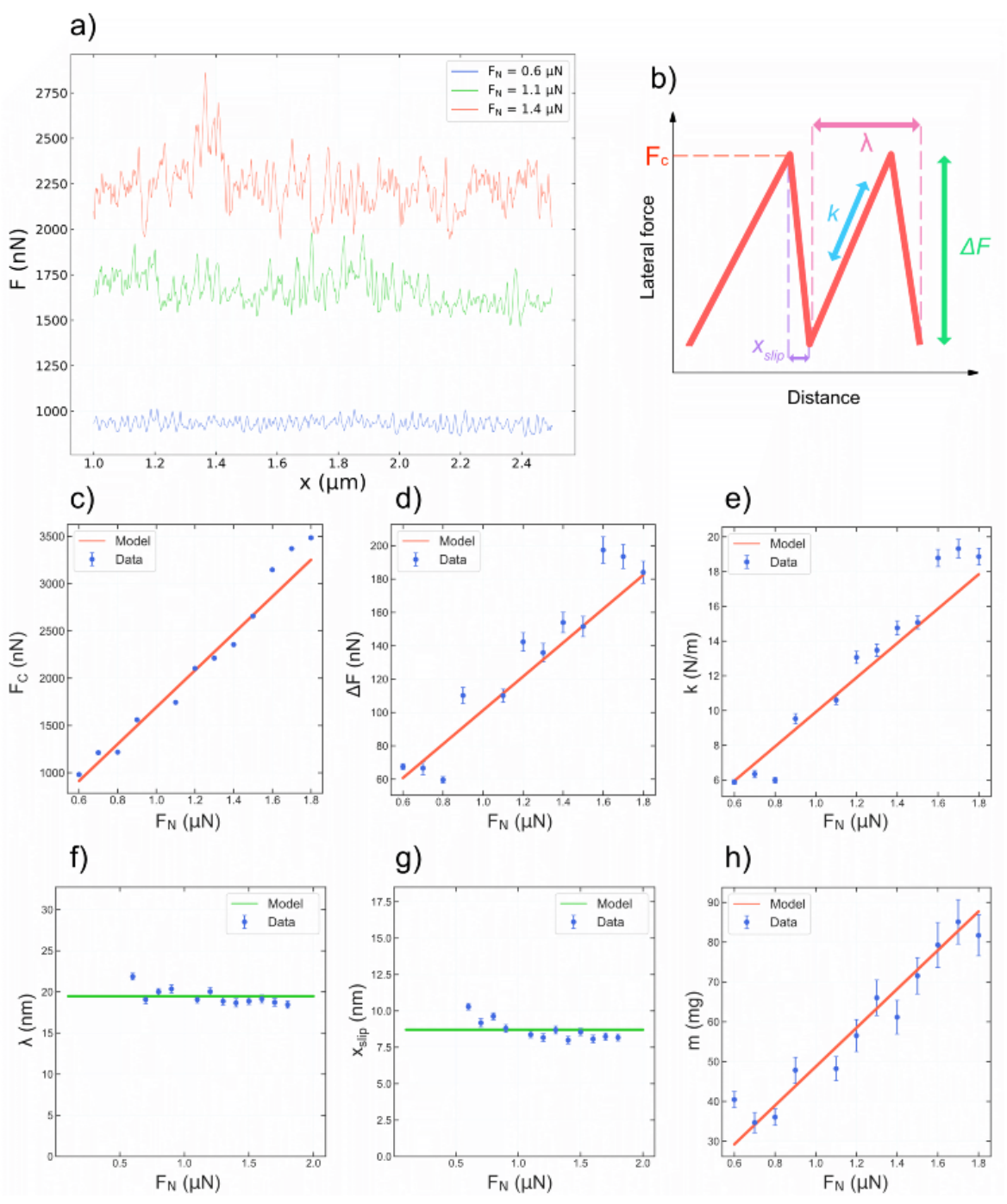}
		\caption{\label{fig:epsart} (a) Lateral force signal recorded while scraping the MoS$_2$ surface with normal forces of 0.6 $\mu$N, 1.1 $\mu$N and 1.4 $\mu$N. (b) Graphical definitions of the critical force $F_c$, force drop $\Delta F$, lateral stiffness $k$, period $\lambda$, and slip length $x_\mathrm{slip}$. (c-h) Load dependence of critical force, force drop, lateral stiffness, period, slip length and effective mass in the stick-slip motion of a diamond tip scratching MoS$_2$ with a scan velocity $v = 1.25$ $\mu$m/s. The red lines in (c-e) are linear fits (with zero intercept) of the experimental data points. The green lines in (f-g) correspond to the average values. The red line in (h) is a linear fit of the data points estimated from the measured values of $\lambda$ and $x_\mathrm{slip}$ using Equation 2.}
	\end{figure*}
	
	For the interpretation of the results in Figure 1 we rely on a simple block and spring model introduced in Ref. \cite{25} and already used in Ref. \cite{26} to describe a diamond tip scratching a silica glass surface. Here, the tip is approximated by a rigid block with mass $m$ elastically driven (with spring constant $k$) on a brittle foundation. Note that $m$ and $k$ must be intended, respectively, as an effective mass and an effective lateral stiffness depending on the coupling between sliding system and substrate established in the contact region. When the spring force reaches a critical value $F_c$, the block starts sliding and, in the present case, also fracturing the material in front of it. The accompanying kinetic friction, $F_k$, causes the process to end within a time $t_\mathrm{slip}$ corresponding to a displacement $x_\mathrm{slip}$ of the spring support (well below the displacement of the tip, which is equal to $\lambda$). In this model, the kinetic friction is simply given by
	\begin{equation}
		F_k=F_c-\frac{\Delta F}{2}.
	\end{equation}
	If $F_k$ is independent of the velocity of the block, the slip time is $t_\mathrm{slip} = \pi/ \omega_0$, where $\omega_0=(k/m)^{1/2}$ is the resonance frequency of the free slider. Since $x_\mathrm{slip} = v t_\mathrm{slip}$, we can estimate the effective mass using the relation
	\begin{equation}
		x_\mathrm{slip} = \pi v \sqrt{\frac{m}{k}}
	\end{equation}
	From the data points in Figure 1(e) and Figure 1(f), $m$ also appears to be proportional to $F_N$, with a coefficient of proportionality of $48 \pm 1$ kg/N, as shown in Figure 2(f). Combined with the experimental relation for $k(F_N)$, we obtain a value of $450 \pm 20$ s$^{-1}$, corresponding to a (load-independent) resonance frequency $f_0 = \omega_0 / (2\pi) = 72 \pm 3$ Hz. This value is a few orders of magnitude below the resonance frequency of the cantilever ($\sim$400 kHz for the free lever and well above when the lever is pinned to the MoS$_2$ surface through the tip). Similarly to the case of a silica glass surface scratched with a nanoindenter setup \cite{26}, it rather corresponds to a low-frequency resonance of the mechanical stage holding the scratching probe. Since $f_0 \gg v x_\mathrm{slip}$ in all cases, the oscillations are strongly overdamped.

	\begin{figure*}
	\includegraphics[width = \textwidth]{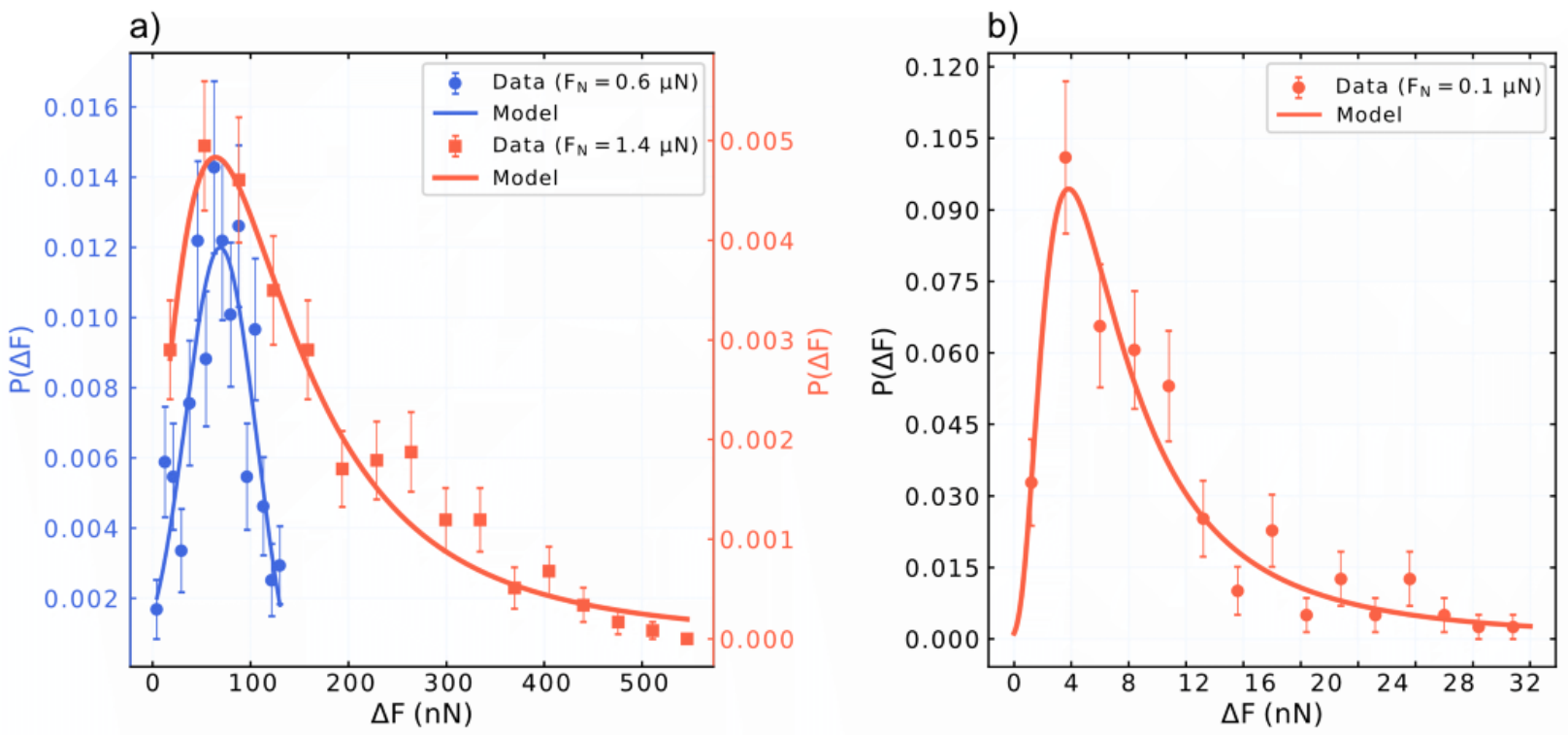}
	\caption{\label{fig:epsart} Statistical distribution of the force drop $\Delta F$ estimated (a) from AFM-based nanoscratch tests on MoS$_2$ with $F_N = 0.6$ $\mu$N (blue dots) and $F_N = 1.4$ $\mu$N (red dots) and (b) from corresponding MD simulations with $F_N = 0.1$ $\mu$N (and a much sharper tip) and three different scan directions. The curves correspond to the best fits according to the Gaussian and GEV distributions introduced in the text.}
	\end{figure*}

	We have also estimated the statistical distribution of the force drop, $P(\Delta F)$, for different values of $F_N$. Two representative curves are shown in Figure 2(a). At a low value of $F_N = 0.6$ $\mu$N, $P(\Delta F)$ is well-fitted by a Gaussian distribution
	\begin{equation}
		P(x) = \exp\left(-\frac{(x-x_0)^2}{2s^2}\right),
	\end{equation}
	with $x_0 = 68 \pm 3$ nN and $s = 34 \pm 4$ nN. However, $P(\Delta F)$ becomes right-skewed when $F_N$ is increased. The situation is reminiscent of frictional stick-slip measurements recently run by Yan \emph{et al.} with flat microtips sliding on sandpaper \cite{27}. In that case, the authors observed that the critical force $F_c$ followed the generalized extreme value (GEV) distribution
	\begin{equation}
		P(x) = \left(1-\frac{c(x-x_0)}{\beta}\right)^{\frac{1}{c}-1} \exp\left[-\left(1-\frac{c(x-x_0)}{\beta}\right)\right]^{\frac{1}{c}},
	\end{equation}
	proposed by Jenkinson to describe sudden changes in meteorological phenomena \cite{28} and later applied to other problems, including earthquake dynamics \cite{29} and the stock market \cite{30}. In the present context, Equation 4 suggests the presence of avalanche effects triggering the slip phase in the complex contact interface formed at sufficiently high values of normal force. With Equation 4 applied to the distribution of $\Delta F$ corresponding to the red points in Figure 2(a), we estimate $x_0 = 90 \pm 5$ nN, $\beta = 82 \pm 5$ nN, and $c = -0.4 \pm 0.1$ when $F_N = 1.4$ $\mu$N. Note that a negative value of $c$ corresponds to a Fréchet distribution \cite{31}.
	
	To conclude this part, we present the surface features resulting from the scratch process, as imaged by AFM. As shown in Figure 3(a-b), the wear groove is decorated by a sequence of exfoliated chips along one of its edges. This is possibly due to the fact that, due to unavoidable asymmetries in the tip shape and orientation, the fracture strength of MoS$_2$ is not reached on both sides at the same time. The chips have a triangular shape, as already shown e.g. in Ref. \cite{10}. Basically, when the fracture strength is reached, the surface is cracked along the scan direction, but the exfoliated chip bends along a straight line toward the closer crystallographic orientation. When the chip reaches a critical size, a second crack is formed. This crack ends back to the scan line and, as a result, the chip remains hinged on one side and usually dangles upward for a few nm (which makes its high resolution imaging very difficult). The average distance between consecutive chips, d, has been also estimated by examining the height profiles extracted along the wear tracks, with the results in Figure 3(e). It increases linearly with $F_N$ as $d = a F_N + b$, with $a = 0.23 \pm 0.09$ m/N and $b = -0.12 \pm 0.07$ $\mu$m. Note that $d$ is comparable to the stick-slip period $\lambda (\sim 20$ nm) only for the lowest values of $F_N$, where a Gaussian distribution of $\Delta F$ is observed. As $F_N$ increases, $d$ becomes one order of magnitude larger than $\lambda$, meaning that several consecutive slip events are required to complete the formation of a chip. Above $F_N = 1.4$ $\mu$N, the chips start to overlap and it is not possible to estimate the distance between them. The cross-section in Figure 3(d) also allows us to estimate 100 nm as an upper limit for the magnitude of the scratch width. With the corresponding value of $F_N = 1.4$ $\mu$N, the normal stress $\sigma$ is in the order of 100 GPa. This is above the value of $\sigma=23$ GPa for the in-plane fracture strength of MoS$_2$ \cite{32}, as it should be.

	\begin{figure*}
		\includegraphics[width = \textwidth]{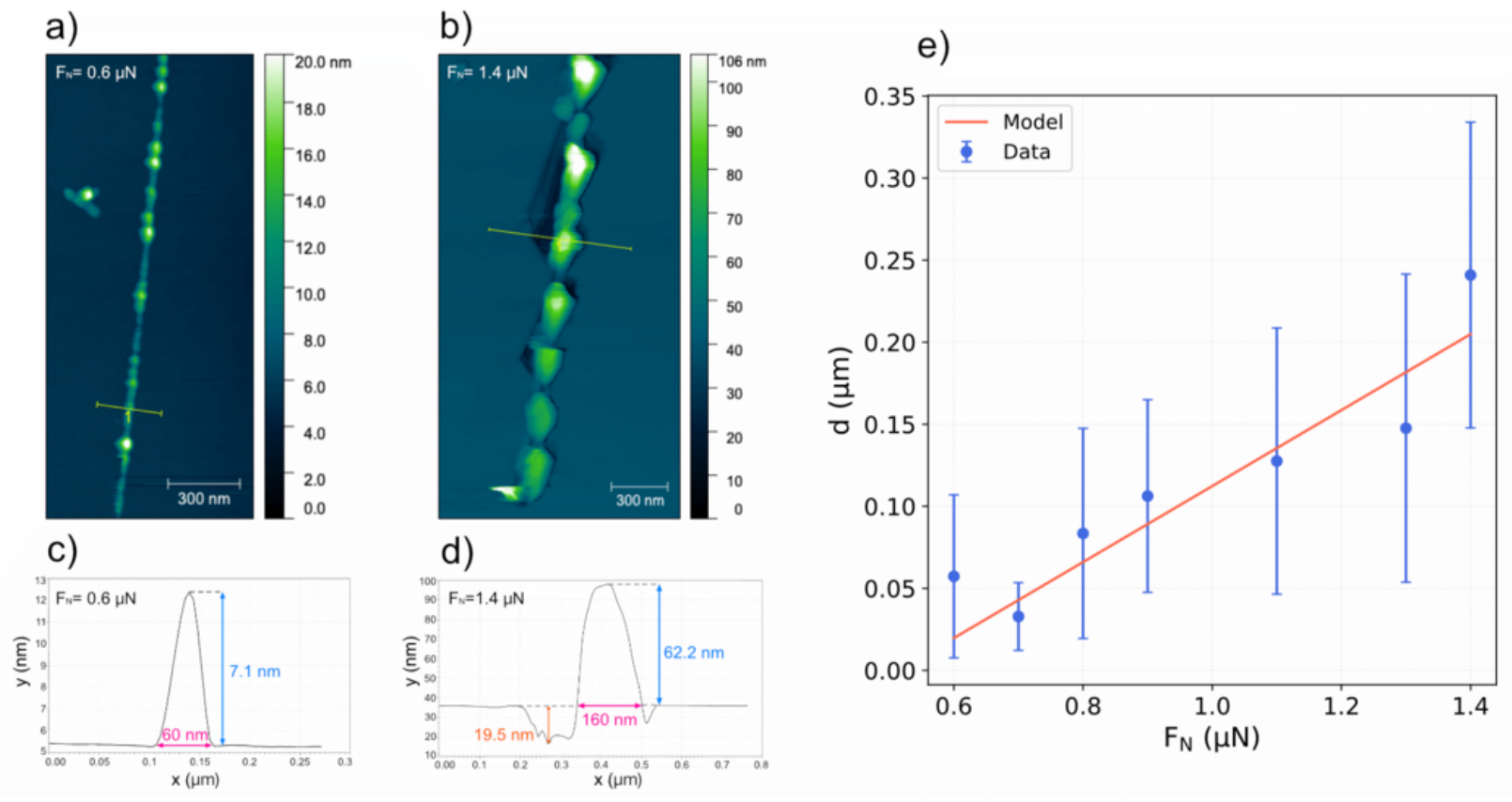}
		\caption{\label{fig:epsart} Chip structures observed after scratching the MoS$_2$ surface with (a) F$_N = 0.6$ $\mu$N and (b) $F_N = 1.4$ $\mu$N. (c, d) Cross-sections corresponding to the yellow lines in (a, b). (e) Average distance between consecutive chips.}
	\end{figure*}
	
	\subsection{Computational Investigation of Stick-Slip and Chip Exfoliation}
	Figure 4(a) shows the time evolution of the lateral force signal, $F$, as estimated by MD simulations when the diamond tip is pulled along the armchair direction of MoS$_2$, and $F_N$ is varied between 100 and 300 nN. The corresponding dependence of the position of the tip apex, $x_\mathrm{tip}$, is shown in Figure 4(b). The stick-slip motion is evident from the sawtooth shape of the $F(t)$ signal and from the corresponding stepwise increase of $x_\mathrm{tip}$ (up to several nm within a few ps). At low loads we can also recognize minor slips with length $< 1$ \AA, and even regions where the tip moves in a continuous way. The simulations have been repeated with the tip pulled at 30$^\circ$ and 15$^\circ$ with respect to the armchair direction (i.e., in the zigzag direction and halfway, see the Supporting Information). The corresponding statistical distributions of the force drops, $\Delta F$, are essentially the same for all directions. These distributions have been averaged leading to the results in Figure 2(b). For all values of $F_N$, $P(\Delta F)$ corresponds to the GEV distribution given by Equation 3. In the case of MD simulations the normal stress $\sigma$ can be estimated from the penetration depth $d$ of the tip, which is plotted in Figure 4(c). With the geometry of our problem, the area indented by the tip is
	\begin{equation}
		A=\frac{\sqrt{3}}{4}\left(\frac{ad}{h}\right)^2,
	\end{equation}
	where $a = 10$ nm is the edge of the tetrahedron, and $h = (2/3)^{1/2} a$ is its height. As a result, $\sigma = F_N / A$ is found to be in the order of 50-80 GPa, which is comparable to the values reached using AFM (with larger values of normal force and contact area).
	
	\begin{figure*}
		\includegraphics[width = \textwidth]{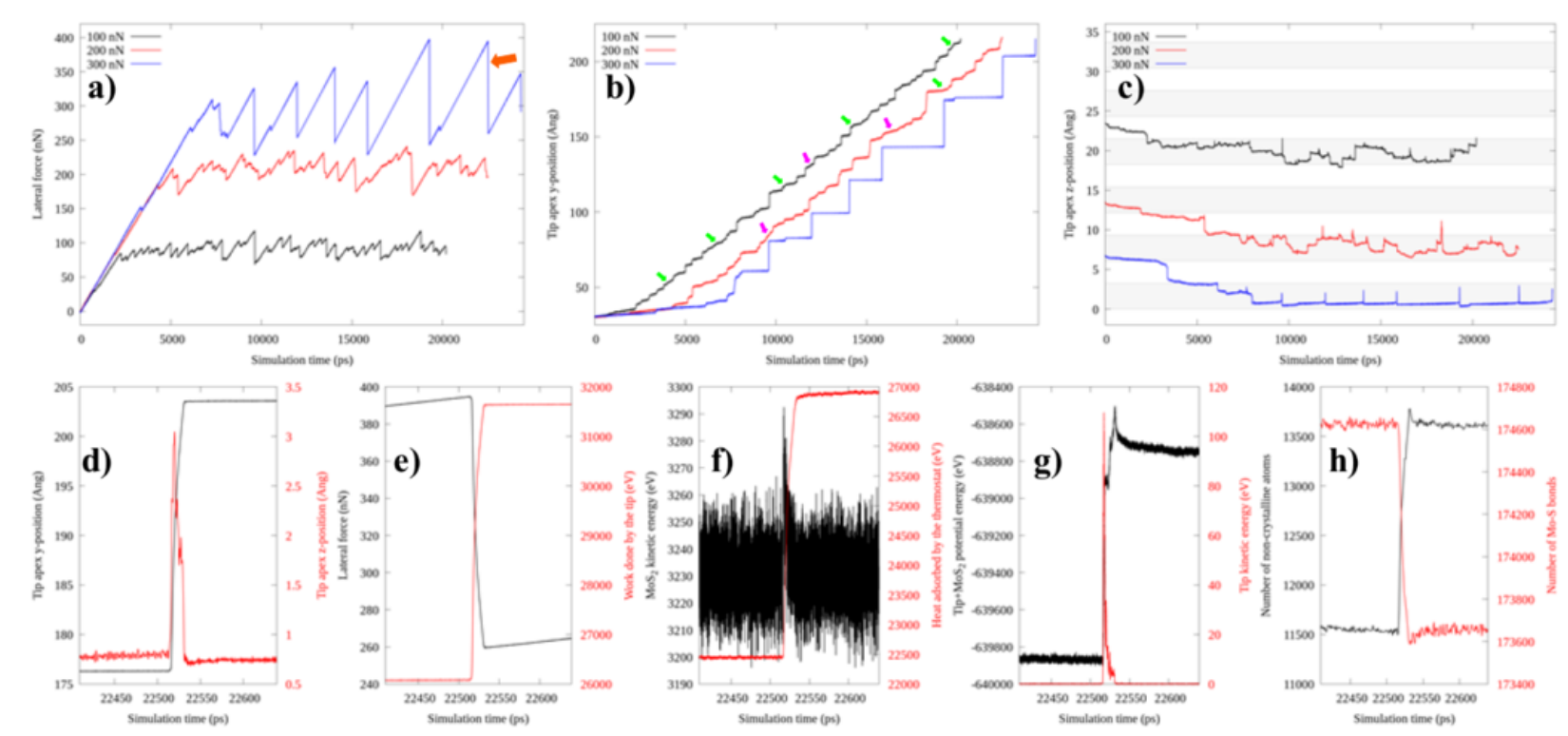}
		\caption{\label{fig:epsart} (a) Time evolution of the spring force $F$ for the three values of $F_N$ used in the MD simulations with the MoS$_2$ surface scratched along the armchair direction. (b, c) Same for the tip position along the scan direction ($x_\mathrm{tip}$, b) and perpendicularly to the MoS$_2$ surface ($z_\mathrm{tip}$, c). In (b), minor sudden slips and regions of continuous tip sliding are marked with green and magenta arrows, respectively. In (c), the initial position of the MoS$_2$ layers is depicted with gray rectangles. (d-h) For the slip highlighted by the orange arrow in (a), profiles of (d) tip displacements, (e) force drop and work done by the tip, (f) MoS$_2$ kinetic energy and heat exchanged with the thermostat, (g) potential energy of the system and kinetic energy of the tip, and (h) number of Mo-S bonds and atoms with a defective neighborhood, are reported as functions of the simulation time.
		}
	\end{figure*}
	
	The time resolution in MD simulations allows us to characterize single slip events much more precisely than in the corresponding AFM-based measurements. In the slip phase highlighted with the orange arrow in Figure 4(a), $t_\mathrm{slip} = 15$ ps. With the spring constant $k = 50$ N/m and the tip mass $m = 1.44$ 10$^{-22}$ kg, this time matches the value of xtip given by Equation 2. The maximum value of the lateral force, $F_c = 395$ nN, is the static friction at the beginning of the slip event. With the force drop $\Delta F = 135$ nN, the kinetic friction acting on the tip while cutting the MoS$_2$ surface is easily estimated from Equation 1 as $F_k = 328$ nN. The work $W$ done by the tip in the slip phase is obtained from the integral of the spring force over the slip distance. As a result, $W = 5.57$ keV. As seen in Figure 4(f-h), a small part ($\sim 0.17$ keV) of this amount of energy is transformed into kinetic energy of both the tip and the substrate during the slip (which are eventually adsorbed as heat by the thermostat). During the slip event, there is also a peak of the potential energy of the system ($\sim 1.36$ keV), which includes contributions from both plastic and elastic deformations of the substrate; the latter are quickly (i.e., in less than 50 ps) relaxed in the subsequent stick phase. Overall, about one fifth of the the energy released by the tip during the slip event is used to increase the internal energy of MoS$_2$ (so, to crack and deform), whilst the remaining part is eventually dissipated through the thermostat (see the Supporting Information for the characterization of all other slip events).
	
	Moreover, MD simulations allow us to identify structural changes in the scratched region of MoS$_2$ at atomistic level, which are also not accessible by AFM. For example, Figure 4(d) reveals that during the slip event the movement of the tip along the scratching direction is accompanied by a retraction of the tip apex of about 2.5 \AA. Figure 5 and the corresponding video in the Supporting Information shows the configuration of our system before and after the slip event highlighted in Figure 4(a) in full details. As the tip cuts the MoS$_2$ substrate, two chips made of two MoS$_2$ layers are formed and exfoliated at either sides of the wear track. The MoS$_2$ surface enters an amorphous state at the edges of the cut, whilst the majority of the chips remain in a crystalline albeit curved state. The profiles in Figure 4(h) allows us to quantify the volume of the material which became amorphous in the selected slip event ($\sim2000$ atoms) and the number of Mo-S bonds broken in the process (about 1000). Last but not least, once the chip reaches a critical size (after a series of few slip events), the wear track suddenly narrows, and a new chip starts to be open, as also noted in Figure 5(b). This opening/closure process is ruled not only by the tip shape, but also by the penetration depth, as suggested by the experimental results in Figure 3(e). This mechanism gives a possible explanation for the GEV distribution of the force drops observed, experimentally, only for sufficiently high loads (Figure 2a). In this case, several stick-slip repetitions are required for the formation of a chip, as in the MD simulations. In each stick-slip event the configuration of the MoS$_2$ surface in contact with the tip-surface contact is qualitatively and quantitatively different, as long as the chip is built up. As a result, it is not surprising that the force drop distribution is distorted with respect to the symmetric (Gaussian) distribution obtained for the lowest value of $F_N$, where the stick-slip period is comparable with the distance between consecutive chips, and the tip-surface contact is similar at each repetition. 
	
	\begin{figure*}
		\includegraphics[width = \textwidth]{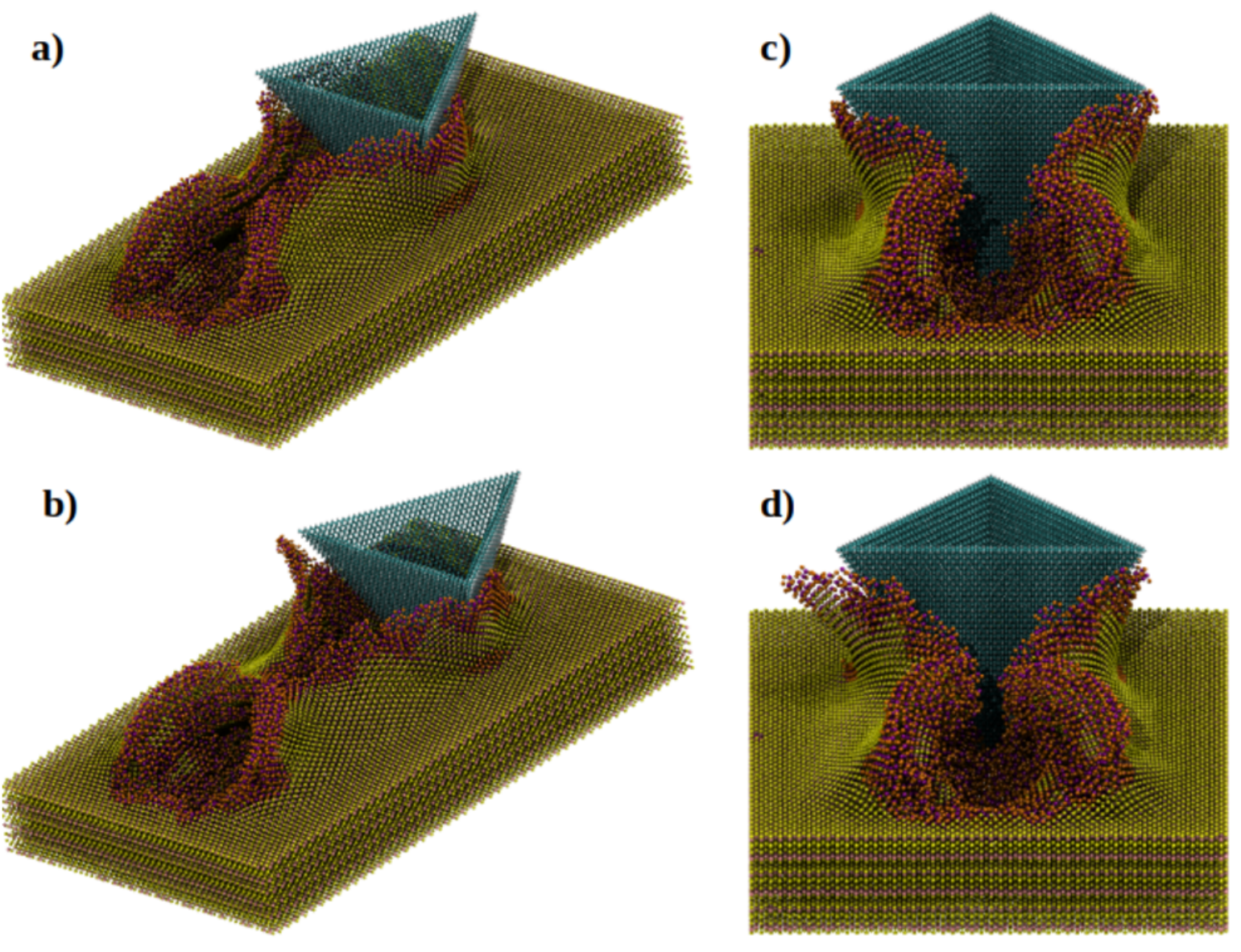}
		\caption{\label{fig:epsart} Snapshots taken before (a,c) and after (b,d) the slip event highlighted with the orange arrow in Figure 5(a). Mo and S atoms in a crystalline(amorphous) configuration are represented in pink (purple) and yellow (orange), respectively. 
		}
	\end{figure*}
	
	\section{Conclusion}
	To sum up, we have investigated the first stages of plowing wear of multilayered MoS$_2$ by a combination of AFM and MD simulations. When a diamond tip is indented for a few nm into the solid surface and elastically driven along it, stick-slip with periodicity well above the in-plane lattice constant of MoS$_2$ is observed. The stick-slip motion is accompanied by progressive exfoliation of the surface, which becomes amorphous all along the cut. The basic mechanism observed with both characterization methods is consistent with a simple “ball-and-spring” model with two distinct values of static and kinetic friction for the stick and the slip phase respectively. Further refinements of the model should take into account the complex evolution of the contact interface suggested by the GEV distribution of the force drop in the slip phase, and also highlighted by MD simulations. In this way, a bridge between fundamental tribology and applied nanotechnology has been established. While our work focuses on MoS$_2$, the mechanisms uncovered here are expected to hold more broadly across layered van der Waals materials, and it can aid the development of better (solid) lubricants, durable coatings, and reliable nanodevices.

		\begin{acknowledgments}
			The support of the Polish National Science Center (NCN) via the Opus Grant No. UMO/2021/43/B/ST5/00705 is gratefully acknowledged by all authors. Computational resources were provided by the e-INFRA CZ project (ID: 90254), supported by the Ministry of Education, Youth and Sports of the Czech Republic.
		\end{acknowledgments}


\begin{thebibliography}{2}
	
\bibitem{1}	Mate, C. M., McClelland, G. M., Erlandsson, R., Chiang, S. (1987). Atomic-scale friction of a tungsten tip on a graphite surface. Physical Review Letters, 59(17), 1942–1945. https://doi.org/10.1103/PhysRevLett.59.1942
\bibitem{2}	Tomanek, D., Zhong, W., Thomas, H. (1991). Calculation of an atomically modulated friction force in atomic-force microscopy. Europhysics Letters, 15(8), 887–892. https://doi.org/10.1209/0295-5075/15/8/014
\bibitem{3}	Lüthi, R., Meyer, E., Bammerlin, M., Howald, L., Haefke, H., Lehmann, T., Loppacher, C., Güntherodt, H.‐J., Gyalog, T., Thomas, H. (1996). Friction on the atomic scale: An ultrahigh-vacuum atomic force microscopy study on ionic crystals. Journal of Vacuum Science and Technology B, 14(2), 1280–1284. https://doi.org/10.1116/1.589081
\bibitem{4}	Liu, Z., Vilhena, J. G., Hinaut, A., Scherb, S., Luo, F., Zhang, J., Glatzel, T., Gnecco, E., Meyer, E. (2023). Moiré-tile manipulation-induced friction switch of graphene on a platinum surface. Nano Letters, 23(10), 4693–4697. https://doi.org/10.1021/acs.nanolett.2c03818
\bibitem{5}	Vilhena, J. G., Pimentel, C., Pedraz, P., Luo, F., Serena, P. A., Pina, C. M., Gnecco, E., Pérez, R. (2016). Atomic-scale sliding friction on graphene in water. ACS Nano, 10(4), 4288– 4293. https://doi.org/10.1021/acsnano.5b07825
\bibitem{6}	Vazirisereshk, M. R., Hasz, K., Carpick, R. W., Martini, A. (2020). Friction anisotropy of MoS$_2$: Effect of tip–sample contact quality. The Journal of Physical Chemistry Letters, 11(16), 6900–6906. https://doi.org/10.1021/acs.jpclett.0c01617
\bibitem{7}	Gnecco, E., Bennewitz, R., Meyer, E. (2002). Abrasive wear on the atomic scale.
Physical Review Letters, 88(21), 215501. https://doi.org/10.1103/PhysRevLett.88.215501
\bibitem{8}	Gosvami, N. N., Filleter, T., Egberts, P., Bennewitz, R. (2010). Microscopic friction studies on metal surfaces. Tribology Letters, 39, 19–24. https://doi.org/10.1007/s11249-009-9508-5
\bibitem{9}	Özoğul, A., Trillitzsch, F., Neumann, C., George, A., Turchanin, A., Gnecco, E. (2020). Plowing-induced nanoexfoliation of mono- and multilayer MoS$_2$ surfaces. Physical Review Materials, 4(3), 033603. https://doi.org/10.1103/PhysRevMaterials.4.033603
\bibitem{10}	Özoğul, A., Gnecco, E., Baykara, M. Z. (2021). Nanolithography-induced exfoliation of layered materials. Applied Surface Science Advances, 6, 100146. https://doi.org/10.1016/j.apsadv.2021.100146
\bibitem{11}	Belak, J., Boercker, D. B., Stowers, I. F. (1993). Simulation of nanometer-scale deformation of metallic and ceramic surfaces. MRS Bulletin, 18(5), 55–60. https://doi.org/10.1557/S088376940004714X
\bibitem{12}	Fang, T. H., Weng, C. I. (2000). Three-dimensional molecular dynamics analysis of processing using a pin tool on the atomic scale. Nanotechnology, 11(3), 148–153. https://doi.org/10.1088/0957-4484/11/3/302
\bibitem{13}	Zhang, J. J., Sun, T., Yan, Y. D., Liang, Y. C., Dong, S. (2008). Molecular dynamics simulation of subsurface deformed layers in AFM-based nanometric cutting process. Applied Surface Science, 254(15), 4774–4779. https://doi.org/10.1016/j.apsusc.2008.01.096
\bibitem{14}	Li, J., Fang, Q., Zhang, L., Liu, Y. (2015). Subsurface damage mechanism of high-speed grinding in single-crystal silicon revealed by atomistic simulations. Applied Surface Science, 324, 464–474. https://doi.org/10.1016/j.apsusc.2014.10.149
\bibitem{15}	Klemenz, A., Pastewka, L., Balakrishna, S. G., Caron, A., Bennewitz, R., Moseler, M. (2014). Atomic-scale mechanisms of friction reduction and wear protection by graphene. Nano Letters, 14(12), 7145–7152. https://doi.org/10.1021/nl5037403
\bibitem{16}	Spalvins, T., Przybyszewski, J. S. (1967). Deposition of sputtered molybdenum disulfide films and friction characteristics of such films in vacuum (NASA Technical Report No. TM-X-794). National Aeronautics and Space Administration.
\bibitem{17}	Martin, J. M., Donnet, C., Le Mogne, T., Epicier, T. (1993). Superlubricity of MoS$_2$. Physical Review B, 48(10), 10583–10586. https://doi.org/10.1103/PhysRevB.48.10583
\bibitem{18}	Peeters, S., Barlini, A., Jain, J., Gosvami, N. N., Righi, M. C. (2022). Adsorption and decomposition of ZDDP on lightweight metallic substrates: Ab initio and experimental insights. Applied Surface Science, 600, 153947. https://doi.org/10.1016/j.apsusc.2022.153947
\bibitem{19}	Nicolini, P., Capozza, R., Restuccia, P., Polcar, T. (2018). Structural ordering of molybdenum disulfide studied via reactive molecular dynamics simulations. ACS Applied Materials and Interfaces, 10(10), 8937–8946. https://doi.org/10.1021/acsami.7b17960
\bibitem{20}	Wei, B., Kong, N., Zhang, J., Li, H., Hong, Z., Zhu, H., Zhuang, Y., Wang, B. (2021). A molecular dynamics study on the tribological behavior of molybdenum disulfide with grain- boundary defects during scratching processes. Friction, 9, 1198–1212. https://doi.org/10.1007/s40544-020-0459-z
\bibitem{21}	Rai, H., Thakur, D., Kumar, D., Pitkar, A., Ye, Z., Balakrishnan, V., Gosvami, N. N. (2022). Spatial variation in nanoscale wear behavior of chemical-vapor-deposited monolayer WS$_2$. Applied	Surface Science, 605, 154783. https://doi.org/10.1016/j.apsusc.2022.154783
\bibitem{22}	Rai, H., Thakur, D., Gadal, A., Ye, Z., Balakrishnan, V., Gosvami, N. N. (2023). Nanoscale friction and wear behavior of a CVD-grown aged WS$_2$ monolayer: The role of wrinkles and surface chemistry. Nanoscale, 15, 10079–10088. https://doi.org/10.1039/D3NR01261A
\bibitem{23}	Huang, C. W., Lai, T. Y., Fang, T. H., Liang, S. W. (2023). Interfacial and tribological characteristics of MoS$_2$ on Ni under nanoindentation and nanoscratch. Physica Status Solidi B, 260(10), 2200555. https://doi.org/10.1002/pssb.202200555
\bibitem{24}	Lantz, M. A., O’Shea, S. J., Welland, M. E., Johnson, K. L. (1997). Atomic-force- microscope study of contact area and friction on NbSe$_2$. Physical Review B, 55(16), 10776–10785. https://doi.org/10.1103/PhysRevB.55.10776
\bibitem{25}	Persson, B. N. J., Spencer, N. D. (1999). Sliding friction: Physical principles and applications. Physics Today, 52(1), 66–68. https://doi.org/10.1063/1.882557
\bibitem{26}	Gnecco, E., Hennig, J., Moayedi, E., Wondraczek, L. (2018). Surface rippling of silica glass surfaces scraped by a diamond indenter. Physical Review Materials, 2(11), 115601. https://doi.org/10.1103/PhysRevMaterials.2.115601
\bibitem{27}	Yan, C., Chen, H. Y., Lai, P. Y., Tong, P. (2023). Statistical laws of stick-slip friction at mesoscale. Nature Communications, 14, 6221. https://doi.org/10.1038/s41467-023-41850-1
\bibitem{28}	Jenkinson, A. F. (1955). The frequency distribution of the annual maximum (or minimum) values of meteorological elements. Quarterly Journal of the Royal Meteorological Society, 81(348), 158–171. https://doi.org/10.1002/qj.49708134804
\bibitem{29}	Pisarenko, V. F., Sornette, A., Sornette, D., Rodkin, M. V. (2008). New approach to the characterization of M max and of the tail of the distribution of earthquake magnitudes. Pure and Applied Geophysics, 165(5), 847–888. https://doi.org/10.1007/s00024-008-0341-9
\bibitem{30}	Bali, T. G. (2003). The generalized extreme value distribution. Economics Letters, 79(3), 423–427. https://doi.org/10.1016/S0165-1765(03)00035-1
\bibitem{31}	Leadbetter, M. R., Lindgren, G., Rootzén, H. (1983). Extremes and related properties of random sequences and processes. Springer. https://doi.org/10.1007/978-1-4612- 5449-2
\bibitem{32}	Bertolazzi, S., Brivio, J., Kis, A. (2011). Stretching and breaking of ultrathin MoS$_2$. ACS Nano, 5(12), 9703–9709. https://doi.org/10.1021/nn203879f
	
	\end{thebibliography}
	\end{document}